\documentclass{aa}
\usepackage{graphicx}
\usepackage{txfonts}
\usepackage{natbib}
\bibpunct{(}{)}{;}{a}{}{,} 

\begin{document}

\title{Study of Globular Cluster \object{M53}: new variables, distance, metallicity}

\author{    D\'ek\'any, I.    \and    Kov\'acs, G.    }

\institute{
    Konkoly Observatory, PO Box 67, H-1525,
    Budapest, Hungary \\ 
    \email {dekany@konkoly.hu} 
}

\date{Received / Accepted }

\titlerunning {Study of \object{M53}}

\abstract
{}
{We study the variable star content of the globular cluster \object{M53} to 
compute the physical parameters of the constituting stars and the 
distance of the cluster.}
{Covering two adjacent seasons in 2007 and 2008, new photometric data 
are gathered for 3048 objects in the field of \object{M53}. By using the OIS 
(Optimal Image Subtraction) method and subsequently TFA (Trend Filtering 
Algorithm), we search for variables in the full sample by using 
Discrete Fourier Transformation and Box-fitting Least Squares method. 
We select variables based on the statistics related to these methods 
combined with visual inspections.} 
{We identified 12 new variables (2 RR~Lyrae stars, 7 short periodic 
stars -- 3 of them are SX~Phe stars -- and 3 long-period variables). 
No eclipsing binaries were found in the present sample. Except for the 
3 (hitherto unknown) Blazhko RR~Lyrae (two RRab and an RRc) stars, 
no multiperiodic variables were found. We showed that after proper period 
shift, the PLC (period--luminosity--color) relation for the first overtone 
RR~Lyrae sample tightly follows the one spanned by the fundamental stars. 
Furthermore, the slope is in agreement with the one derived from other 
clusters. Based on the earlier Baade-Wesselink calibration of the PLC 
relations, the derived reddening-free distance modulus of \object{M53} is 
$16.31\pm0.04$ mag, corresponding to a distance modulus of $18.5$~mag 
for the Large Magellanic Cloud. From the Fourier parameters of the RRab stars we obtained an 
average iron abundance of $-1.58\pm 0.03$~(error of the mean). This is 
$\sim 0.5$~dex higher than the overall abundance of the giants as given 
in the literature and derived in this paper from the three-color 
photometry of giants. We suspect that the source of this discrepancy 
(observable also in other, low-metallicity clusters) is the want of 
sufficient number of low-metallicity objects in the calibrating sample 
of the Fourier method.}
{}


\maketitle

\section{Introduction}

Among the oldest and most metal-poor Galactic globular 
clusters, the outer halo cluster \object{M53} (\object{NGC\,5024}, $\alpha=13^{\rm h} 12^{\rm m} 55\fs3$, 
$\delta=+18\degr 10\arcmin 9\arcsec$) is the second most abundant 
in variable stars after \object{M15} (\citealt{cvsgc}, see also \citealt{acs2009} 
for current age determinations). 
The search for its RR~Lyrae content dates back to \citet{shapley1920}, 
and yielded a total of 60 RR~Lyrae stars until now. 
A thorough review of discoveries is given in the paper of \citet{kop2000}, 
the only CCD time-series photometric study of \object{M53} RR~Lyrae stars prior to this work. 
This cluster is also very abundant in blue straggler stars (BSSs). Among the 
almost 200 BSSs, 8 were proven to be SX~Phe variables \citep{jeon2003}. 
We also note that \citet{beccari2008} showed that the BSSs follow a bimodal 
distribution, implying a binary rate of $10\%$, but no eclipsing binaries have been found so far.
As for the most extensive color-magnitude study of \object{M53}, we refer to 
\citet{rey1998}. The cluster is located at a very high Galactic latitude of 
$b=79\fdg76$. Therefore, its field contamination and interstellar reddening 
is expected to be negligible. This is confirmed by \citet{zinn1985}, reporting 
E$(B-V)=0.0$, and by \citet{schlegel}, who give a reddening value of E$(B-V)=0.02$. 
Despite its high variable content and its favorable position, 
no comprehensive wide-field CCD time-series photometric study has been published 
for this cluster. 
Among the two previous works of similar type, \citet{kop2000} observed only 
the central part of the cluster, while \citet{jeon2003} limited themselves 
to the investigation of SX~Phe stars. The present study, based on two-color 
photometry covering most of the cluster, intends to improve the time 
and spatial coverage of this cluster.

To exploit the entirety of our photometric data-base and detect 
as many variables as possible, we employ the widely used method of 
Optimal Image Subtraction (OIS) developed by \citet{al98} and 
improved by \citet{alard2000}, to obtain optimum relative light 
curves. Then, in a post-processing phase, we filter out temporal 
trends by using the Trend Filtering Algorithm (TFA) of \citet{kbn2005}. 
This latter method has been extensively used during the past several 
years by the HATNet Project\footnote{Hungarian-made Automated 
Telescope Network, see \\ \texttt{http://cfa-www.harvard.edu/$\sim$gbakos/HAT}} 
in search for transiting extrasolar planets 
and also employed successfully in general variable search 
\citep{kob2008,szkw2009}. The main purpose of this paper is 
to search for variables in the full dataset and not limit the 
investigation to the well-known regions of variability (i.e., the RR~Lyrae 
strip, the blue straggler and the upper red giant branch regions of the
Hertzspring-Russell diagram). 
With the aid of the reduction and post-processing methods mentioned 
above, we can increase photometric precision and attempt to reach 
the detection limit set by the photon statistics.

In Sect.~\ref{method} we give details on the observations and data 
processing, and show examples of the detections resulting from the 
methods applied.
The next three sections (Sects.~\ref{rrlyr}--\ref{lpv}) are devoted 
to the more detailed analysis of the specific types of variables, 
most importantly to that of the RR~Lyrae stars. Our attention will be 
focused mostly on the period--luminosity--color relation and the metallicity 
issue. A brief discussion and conclusions are given in Sect.~\ref{conclusions}.

\section{Observations, data reduction, method of analysis}
\label{method}

\subsection{Observations and data reduction}
\label{reduction}

We performed time-series photometric observations of \object{M53} through 
$V$ and $I$ filters of the Johnson and Kron-Cousins system, 
respectively, with the $60/90/180$\,cm Schmidt Camera of the Konkoly 
Observatory located at Piszk\'estet\H{o}. The telescope is 
equipped with a Photometrics AT 200 type CCD of 9-$\mu$ pixels 
in a $1536 \times 1024$ array. The point spread function (PSF) is 
slightly undersampled with a scale of $1.026$ arcsec/pixel. The observations were 
carried out between March 25, 2007 and May 28, 2008 in two seasons, 
on a total number of 26 nights. The observations have a full time 
span of $430$~days. Exposure times varied between $550$ and $750$ seconds 
in $V$ and $350$ and $540$ seconds in $I$ depending on 
atmospheric conditions. A total number of $340$ and $230$ frames have been  
collected in the $V$ and $I$ bands, respectively.
While for the time series analysis we use only the $V$-band data, 
for standard magnitude transformations, variable star classification, 
and for deriving cluster distance and metallicity, we also utilize 
the $I$-band data.

The processing of the images (bias, dark, and flat-field corrections) and 
the alignment of all frames into a common pixel reference system were 
performed by standard \textsc{iraf}\footnote{\textsc{IRAF} is distributed 
by the National Optical Astronomy Observatory, which is operated by the 
Association of Universities for Research in Astronomy, Inc., under 
cooperative agreement with the National Science Foundation} packages.
The joint field of view of all observations was somewhat truncated compared 
to the full detector area due to technical reasons. Only this 
$13\farcm5 \times 14\farcm5$ common area (still covering a fairly large part
of the cluster by spanning to $\sim 0.7$ tidal radius, see the catalog of 
\citealt{harris1996}, revised in February 2003), centered approximately 
on the cluster core, was subjected for further reduction. 

We calibrated the astrometric transformations between pixels and celestial 
coordinates by using 30 bright stars from the second edition of the Guide 
Star Catalog \citep[version 2.3.2,][]{gsc2} uniformly distributed around 
the cluster center and located sufficiently well outside the cluster core. 
The standard deviations in the residuals of the coordinate mapping were 
$0\farcs148$ and $0\farcs163$ in right ascension and declination, respectively. 

In order to deal with heavy crowding and to probe the regions close to the 
center of the cluster, we applied the OIS method of \citet{alard2000}.
For this method one has to prepare an accurate reference frame (RF) 
in each band which will be subtracted from all other frames after 
the appropriate scaling of the PSF of the stellar objects. 
The success of the method is highly sensitive to the 
quality of the RFs, therefore we carefully selected the best images 
with the lowest background, best seeing, and most regular PSFs and stacked 
them into one single image in order to increase their signal-to-noise 
ratio. Before the subtraction from each frame, the RF was convolved 
with a kernel function, which was determined by using the representative PSFs 
of 225 stars uniformly distributed on the image. The kernel functions 
were allowed to have second order spatial variations to account for a 
varying PSF across the image. The convolved RF, being the best approximation 
of the processed frame in a least-squares sense, was subtracted from 
the processed frame. The method yielded a set of residual images 
consisting of, in principle (but not in practice), the true light variations 
plus noise. For all procedures of image convolution and subtraction we used 
the \textsc{isis 2.2} package of \citet{alard2000}.

We applied the \textsc{daophot} PSF-fitting software package \citep{daophot} 
on the RFs to identify stellar objects and to obtain their instrumental 
magnitudes. We found 3048 stars above the $10\,\sigma$ level. Linear 
transformation equations of these instrumental magnitudes into standard 
ones were established by using 110 photometric standard stars of 
\citet{standards} around the cluster center.

We measured differential flux variations on the residual frames by aperture 
photometry at all stellar positions obtained from the RFs, using a custom 
\textsc{IRAF} routine. The sizes of the apertures were variable in order to 
conform to the FWHMs of the PSFs on the original frames, changing in time 
due to seeing variations.
To diagnose if some faint variables might have been remained undetected on 
the RFs due to crowding but had light variations emerged on the residuals, we also
employed the detection algorithms of {\sc isis 2.2} in two ways. First, all
sources of significant variability were checked for the case if light variation 
comes from a yet undetected faint star falling within the same aperture. In these 
cases we performed its photometry using a new, accurately recentered aperture. Secondly, we 
checked if there were any remaining faint variables that had neither been detected on the 
RFs nor fell within any aperture (we note that we did not find any additional 
variables by this latter test). All light curves obtained in this way, expressed 
in linear flux units, were then further processed with the aid of the Trend 
Filtering Algorithm (TFA, see Sect.~\ref{tfa} for further details).

As the last step of the reduction, to convert differential fluxes of an object 
into magnitudes, one has to know the flux $f_{\mathrm{RF}}$ of the source on the 
RF in each band which has been subtracted from each processed frame 
\citep[see e.g.,][]{woz2000}.
This additional information sets the zero point of the logarithmic transformation 
$$\Delta m_i = -2.5\log(f_{\mathrm{RF}} + \Delta f_i)$$
between the $\Delta m_i$ differential magnitude and $\Delta f_i$ differential 
flux of the object (derived by the OIS method) on the $i$-th frame. The 
$f_{\mathrm{RF}}$ value has been measured by PSF photometry for each object 
as described above. Clearly, a flux-to-magnitude transformation with an 
imprecise $f_{\mathrm{RF}}$ value yields an incorrect amplitude of the light 
variation and can significantly alter the magnitude averages as well, 
but its actual effect on these errors is specific to each object and 
depends on the brightness and the shape of the light variation. 
Therefore, the flux curves of variable stars were transformed into magnitudes 
only if their (already standardized) photometry on the RFs were reliable enough, 
which has been decided based on the formal errors of the PSF photometry. 
Generally, objects were transformed 
into magnitudes if the formal error of their $(V-I)_{\rm RF}$ was below 
$0.03$\,mag (see Sect.~\ref{plc} for details on this cut in the case of 
RR~Lyrae stars)\footnote{ Photometric time-series of the variables detected 
in this work and the Fourier decompositions of the RR Lyrae stars with 
magnitude-transformed light curves in $V$ color are available electronically 
online at CDS.}.

\subsection{Post-processing with TFA}
\label{tfa}

The time-series data-base obtained above is subjected to a subsequent 
post-processing phase, when we filter out various trends/systematics 
due to, e.g., imperfect reduction, lack of correction to position- 
and time-dependent extinction, anomalies in the convolution procedure 
prior to image subtraction, etc.
The method is described in detail by \citet{kbn2005} and also 
summarized recently by \citet{bak2009} and \citet{szkw2009}. 
Here we only briefly note that the method is based on the idea of 
correcting elements of the systematic variation in the target time 
series by using the light curves of many other objects, available in the CCD 
frame. In the course of signal search, we have no information 
on the temporal content of the time-series. Therefore, we assume that 
the target is trend- and noise-dominated. If it is the case, then 
the method suppresses the trend and gives rise to the signal to appear 
in some time-frequency transformation of the detrended signal. 
Here, on the expense of the suppression of the systematics, we also 
deform the true signal at some level. This effect is cured in the 
second step of filtering, when we reconstruct the signal by using 
a full time-series model that includes both the systematics and the 
signal with the period obtained in the first step. An extension of 
the method to multiperiodic signals is given in \citet{kob2008}. 
In this method (adopted also in here) one fits the TFA 
template and the signal (represented by a Fourier series) 
simultaneously, thereby avoiding iteration by incomplete time-series 
model representations.

\begin{figure}
    \centering
    \includegraphics[width=88mm]{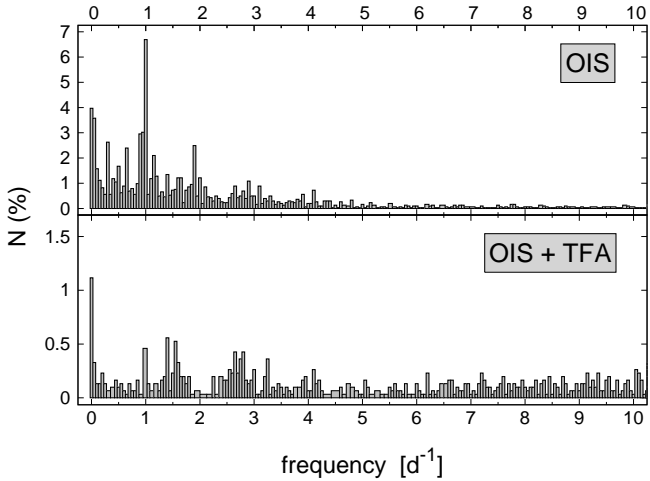}
    \caption{
Histogram of the peak frequencies of the Fourier (DFT) spectra for 
the 3048 stars analyzed. Upper panel shows the frequency distribution 
of data obtained by aperture photometry on the residual frames of the 
OIS method. Lower panel shows the effect of TFA on these data.
        }
    \label{histogram}
\end{figure}

Special attention is to be paid here to the effects related to the 
number of data points. In ideal case, the necessary number of 
TFA templates to handle all systematics is considerably lower than 
the number of data points constituting the time-series. Unfortunately, 
we do not have a good method to select a ``representative'' small 
set of templates, therefore we use a large set that includes both 
``useful'' and ``useless'' (e.g., pure noise) template time-series. 
Attempts have been made to determine an optimum number of templates 
but it seems that it is hard to settle at a value lower than a few 
hundred \citep{szkw2009}. Since in the present case we have 
only $340$ data points per time-series, we have to limit the number 
of templates considerably (i.e., about half of the number of the data points) 
to avoid overfitting and sudden increase of the false alarm probability 
(FAP). This latter effect is thoroughly tested in each interesting 
case by using multiple template runs, which are capable of reducing 
FAP by several factors \citep{kob2007}. To avoid significant 
overfitting, we settled on the lowest TFA template numbers that yielded 
the most significant detection together with the lowest unbiased scatter 
around the signal. This optimum template number varied between 
$60$ and $140$, the latter being an upper limit set 
by us due to the low number of available data points.

In general, we selected the TFA template sets from the $\sim1000$ 
brightest stars in the sample. The faintest stars in the template 
sets did not exceed $\sim18$~mag in $V$ band. The templates 
cover the frame in a quasi-uniform manner 
\citep[see][for more details on the distribution of templates]{kbn2005}. 

The significance of the Fourier or BLS components \citep[see][]{kzm2002} 
was deduced by checking the SNR of the frequency spectra (the ratio of the 
amplitude of the highest peak to the standard deviation of the spectrum). 
Statistical tests \citep[similar to the ones performed in our earlier 
papers, e.g., in][]{nk2006} have led to the conclusion that for 
Gaussian signals with ${\rm SNR}>6.7$, the FAP due to uncorrelated 
random noise is less than $1\%$. Considering that the processed 
signals always have some colored noise (either physical or instrumental) 
we selected variables well above this SNR value (usually with SNR 
exceeding $\sim 8$). 

In demonstrating the ability of signal detection, first we compare 
the distributions of the peak frequency components obtained by 
OIS and by the subsequent application of TFA on the OIS time 
series. Figure~1 shows that already the OIS data are reasonably 
free of periodic systematics. Actually, when compared with 
earlier similar diagrams derived on data of aperture photometry 
\citep[e.g.,][]{kbn2005,szkw2009} we see that the current data 
are much less dominated by the customary $1$\,d$^{-1}$ systematics.
We see in the bottom panel that TFA has successfully filtered out 
the daily trends. In the final dataset the frequency distribution 
becomes nearly flat, with a slight surplus due to the RR~Lyrae stars 
and the long-term irregular light variations of red giants. 

\begin{figure}
    \centering
    \includegraphics[width=88mm]{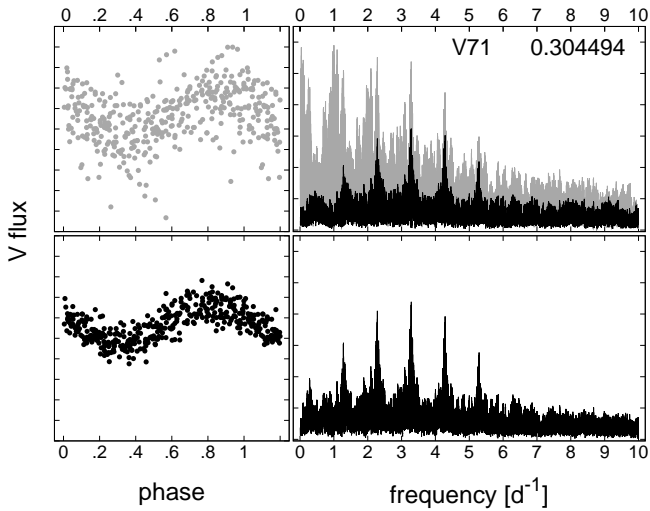}
    \caption{
Detection of the RRc variable V71. Top left: folded light curve 
(with the period detected by TFA) of the OIS data; top right: 
DFT of the OIS data (light shade) 
and that of the TFAd data (dark shade); bottom left: 
folded light curve of the TFA-reconstructed data; 
bottom right: DFT of the TFA-reconstructed data. Star 
identification name and period (in days) are shown in the 
upper right corner. Please check note in text on the scatter 
of the TFA-filtered light curve.
        }
    \label{demo-rrc}
\end{figure}

\begin{figure}
\centering
\includegraphics[width=88mm]{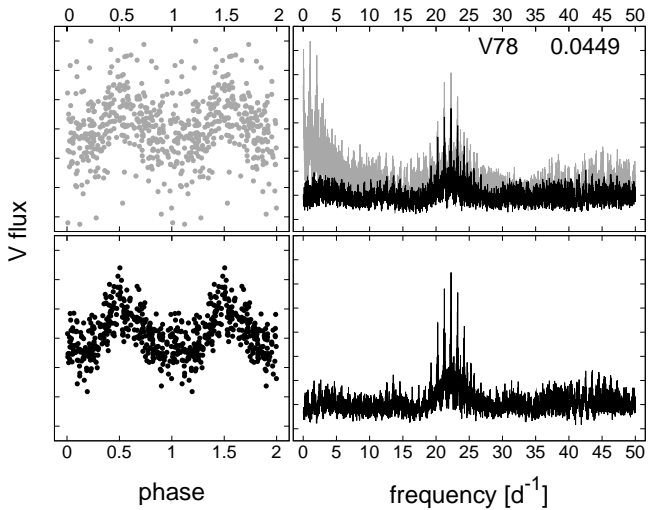}
\caption{
Detection of the SX~Phe star V78.
Notation is the same as in Fig.~\ref{demo-rrc}.
}
\label{demo-spv}
\end{figure}

In checking the effect of TFA in more detail, we plotted the DFT 
(Discrete Fourier Transformation) spectra 
and the folded light curves for the OIS and OIS$+$TFA data in the case of an RRc 
(V71, Fig.~\ref{demo-rrc}) and an SX~Phe star (V78, Fig.~\ref{demo-spv}). 
The effect of the signal reconstruction is large 
in both cases. The true variability of V71 would have been significantly more 
troublesome to decipher from the forest of high peaks in the original 
(OIS) data than in the clean, single-component spectrum of the 
TFA-filtered time-series. For V78 the situation is better because of the high frequency of 
the pulsation. Nevertheless, even in this case, the highest peak is near 
$1$\,d$^{-1}$ (and its aliases), therefore, an automatic search that only looks 
for the highest peak in the spectra, would not find this variable in the 
original (OIS) data. The improvement in the quality of the folded 
light curve derived from the reconstructed data is also substantial. We should 
note however that the small scatter seen is somewhat biased, because of 
the large number of TFA templates used.
The effect of the bias in the 
residual scatter is common to all data regressions. The fit introduces 
correlation among the original array elements that leads to smaller observed 
scatter as it were expected if only the deterministic (model-predicted) part 
of the signal would have been fitted. In general, when least-squares 
fitting a data array of $N$ data points with $N_p$ parameters (larger 
than exact model representation required) the residual RMS will be 
$\sqrt{(N-N_p)/N}$-times of the value expected from the noise 
level in the original data (see also \citealt{szkw2009} for further details 
on this subject). For this reason, the unbiased standard deviations 
of the reconstructed light curves are $10\%$ and $30\%$ higher due to the $60$ and $140$ 
templates used for V71 and V78, respectively. The different template 
numbers resulted from the multi-template tests mentioned earlier in this 
section, in the description of the method of analysis. 

\begin{figure}
\centering
\includegraphics[width=8.0cm]{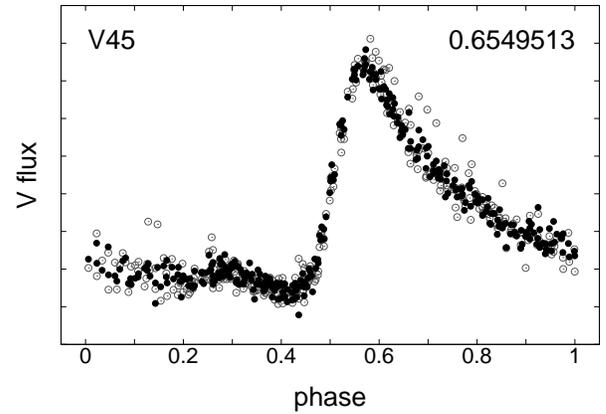}
\caption{
Signal reconstruction in the case of the RRab star 
V45. Open circles denote the OIS fluxes while 
filled ones show the TFA-processed values.
Please check note in text on the scatter of the 
filtered light curve.
}
\label{demo-rrab}
\end{figure}

In a final example, in Fig.~\ref{demo-rrab} we show the signal 
reconstruction capability of TFA. We used an $8$th order Fourier 
series with $144$ TFA templates in the reconstruction 
\citep[see][]{kob2008}. Inspection of the DFT spectrum of the OIS 
light curve of this star shows that it is not influenced by periodic systematics 
of type integer d$^{-1}$. Therefore, the improvement introduced by TFA is 
attributed to filtering out some transient signals. 
We found transient systematics also in other 
variables, similarly to the ones reported by \citet{szkw2009}.
The unbiased standard deviation of the residuals around the 
best-fitting Fourier-sum dropped down by $14\%$, which is a good
improvement over the original data.

\section{RR Lyrae stars}
\label{rrlyr}

Our prime interest in this paper is the study of the RR~Lyrae stars, 
since they are in large number in the cluster and they are the ones 
that have fairly well established theoretical and observational 
understanding (e.g., variable and mode identification, computation 
of the physical parameters from observed ones, etc). We focus on 
the distance and on the metallicity (i.e., [Fe/H]) as the two 
most important parameters derivable from the available empirical 
relations.

\subsection{General description}
\label{gendesc}

We identified altogether $54$ RR~Lyrae stars in the cluster. 
Except for the two RRc stars V71 and V72, all others were 
known previously \citep[see the Catalog of Variable Stars in 
Globular Clusters, hereafter CVSGC,][]{cvsgc}. We note that 
the variability status of V34, which had been questioned 
previously by \citet{vgend47} was clearly confirmed by our data. 
Due to the application of OIS and TFA, 
many of the variables ended up with light curves suitable for 
further analysis. We give the full list of variables together 
with their basic observed parameters in Table~\ref{rrlprop}. Please 
note the very high SNR attached to each detection even for the 
variables sitting near the center of the cluster.

The accuracy of the periods of the RR~Lyrae stars has been estimated 
in the following way. First we computed the Fourier decompositions 
from the TFA-reconstructed light curves. From these decompositions 
synthetic light curves were generated and Gaussian noise was added 
with standard deviations obtained from the unbiased estimates of the 
residual scatter between the synthetic light curves and the TFA 
reconstructions. The generated light curves were subjected to the 
standard Fourier frequency analysis as if they were real observed 
light curves. The periods obtained from each realization were stored 
and after completing 100 independent simulations for each object, the 
standard deviations of those period values were computed. The errors 
shown in Table~\ref{rrlprop} are these standard deviations. 

The periods obtained in this study are generally in good agreement 
with previously determined values, except for the three RRc stars 
V44, V47, and V55, for which previous values are various aliases of 
the true periods. In the case of other first overtone stars the 
agreement between our and earlier published values are better 
than $0.0004\,{\rm d}^{-1}$ with a typical difference of about 
$0.00003\,{\rm d}^{-1}$. As for RRab stars, deviations are generally 
below $0.00002\,{\rm d}^{-1}$, with a few higher values not exceeding 
$0.00012\,{\rm d}^{-1}$. 

\begin{figure*}
\centering
\includegraphics[width=17cm]{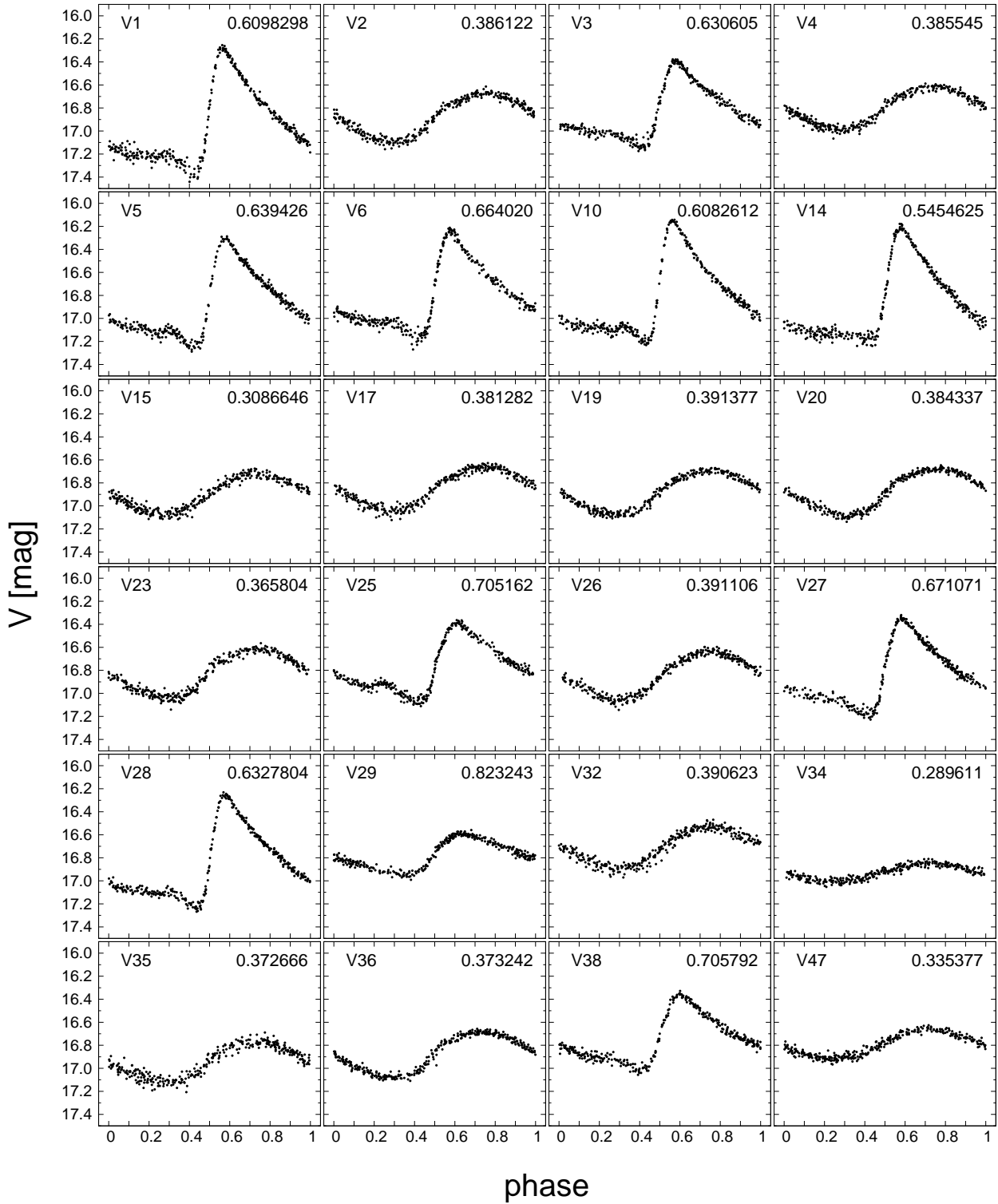}
\caption{
Standard $V$ light curves of the 24 RR~Lyrae stars for which the 
flux-to-magnitude conversion was performed. CVSGC numbers and periods 
in days are given in each case (see also Table~\ref{rrlprop} for 
additional data). All light curves were reconstructed by TFA.
}
\label{rrlyrlc}
\end{figure*}

Figures~\ref{rrlyrlc} and \ref{rrlyrfc} 
show the TFA-reconstructed and folded light curves. Standardizing 
the photometry for the highly crowded $30$ objects shown in 
Fig.~\ref{rrlyrfc} was not reliable enough, therefore, we left the light 
curves in relative flux units. Unfortunately, these objects can only be 
partially utilized here due to the lack of empirical relations 
derived on fluxes and the absence of existing photometry with a 
higher resolution CCD \citep[exceptions are V7, V8, V9, V33, and 
V37 that have good-quality light curves in $V$ magnitude from][]{kop2000}. 
We note that among the previously known RR~Lyrae stars missing from 
our sample, V52 and V53 were merged with each other, V61 was merged 
with the long period variable V49, and the rest (V12, V13, V21, V30, 
and V48) were outside the field of view.

\subsubsection{Blazhko-stars}
\begin{figure*}
\centering
\includegraphics[width=17cm]{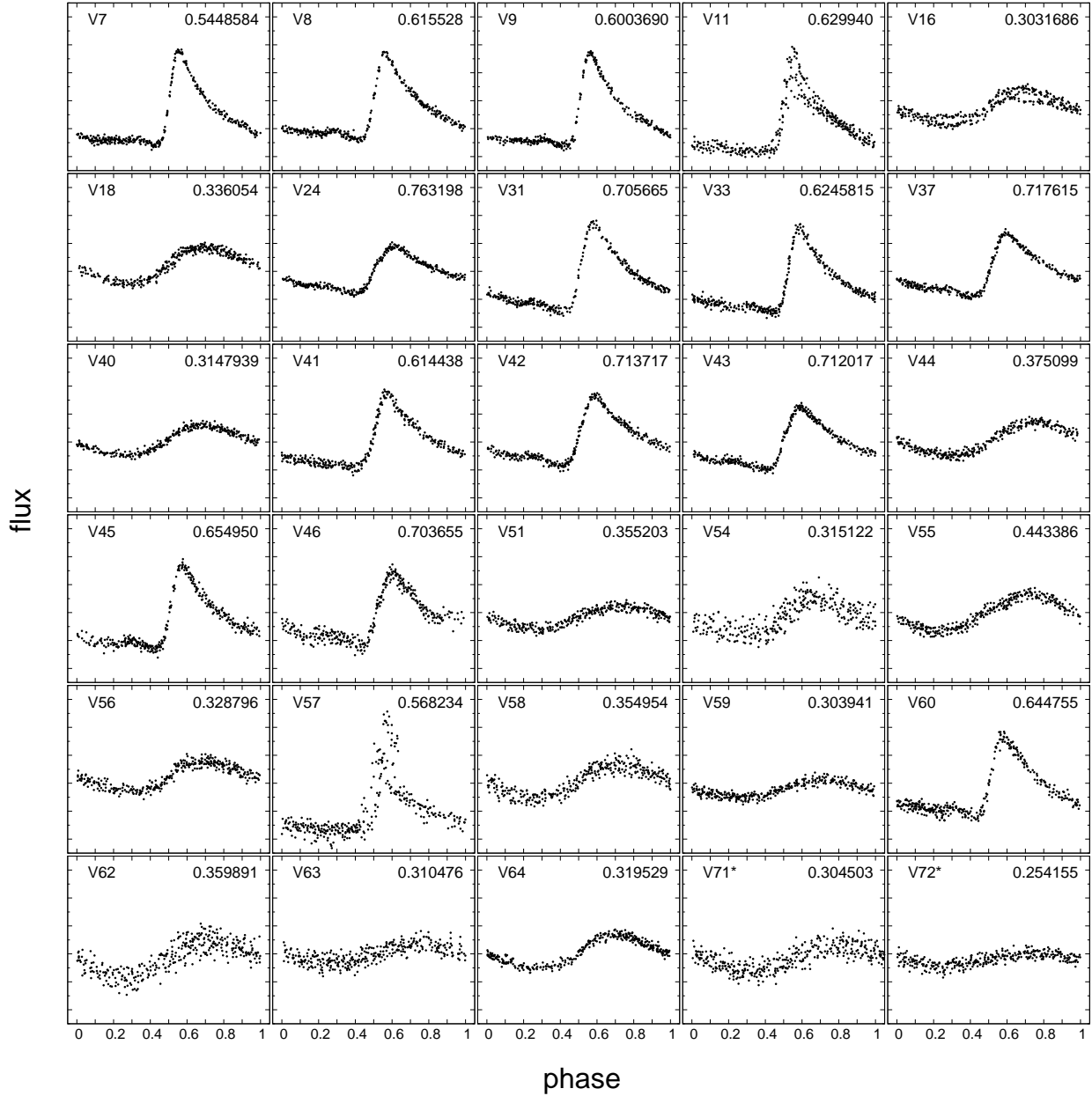}
\caption{
Light curves expressed in arbitrary differential flux units of the 30 RR~Lyrae stars 
that were not transformed into magnitude values due to the large zero-point errors associated 
with crowding. All plots are shown on the same ordinate scale. CVSGC numbers and periods in days 
are given in each case (see also Table~\ref{rrlprop} for additional data). All light curves were 
reconstructed by TFA.
}
\label{rrlyrfc}
\end{figure*}

We investigated possible multiperiodicity by performing successive 
prewhitening on all $3048$ objects, including RR~Lyrae stars. 
Surprisingly, from the whole sample we found only the three Blazhko 
stars given in Tables~\ref{rrlprop} and \ref{blazhko}, in spite of the relatively large 
scatter visible in several stars (see Fig.~\ref{rrlyrfc}). We note 
that the detection of the Blazhko behavior in the three stars 
is firm. The modulation is clearly observable 
as side-lobe frequencies with uniform $f_m$ frequency separations after 
prewhitening the main frequency and its harmonics. For V11, significant 
side lobe frequencies could be identified for up to the 7th harmonic 
of $f_0$ at $nf_0+f_{\rm m}$, except for $n=2$, where we found 
$2f_0-f_{\rm m}$. In the case of the RRc star V16 we found $f_0+f_{\rm m}$ 
and $2f_0+f_{\rm m}$. The phase modulations in both stars are very little, 
unlike to that of the third Blazhko-star, V57, whose modulation is very 
large both in phase and in amplitude. We detected $f_0+f_{\rm m}$ and 
$2f_0+f_{\rm m}$ in the case of V57. The light curves of the 
Blazhko stars folded with the pulsation periods are shown in Fig.~\ref{rrlyrfc}. 
The light curves were reconstructed by simultaneously fitting the 
template and the full modulated signal (see Sect.~\ref{tfa}).

\begin{table*}
\caption{Properties of the RR~Lyrae stars}
\label{rrlprop}
\centering
\begin{tabular}{l l c c r l c c c}     
\hline\hline
      ID    &   Type  & $\alpha$ [hms] & $\delta$ [dms] &$d$ [\arcsec]&   $P$ [d]     &   SNR     &    $V$     &   $V-I$  \\
\hline
      V1    &   RRab  &   13:12:56.3   &   18:07:13.8   &   176.8   &   0.6098298(9)  &   14.3    &   16.970   &   0.579  \\
      V2    &   RRc   &   13:12:50.3   &   18:07:00.8   &   202.0   &   0.386122(1)   &   22.2    &   16.879   &   0.402  \\
      V3    &   RRab  &   13:12:51.4   &   18:07:45.5   &   154.6   &   0.630605(1)   &   17.8    &   16.848   &   0.447  \\
      V4    &   RRc   &   13:12:43.9   &   18:07:26.4   &   230.3   &   0.385545(1)   &   26.0    &   16.800   &   0.383  \\
      V5    &   RRab  &   13:12:39.1   &   18:05:42.6   &   353.3   &   0.639426(1)   &   23.2    &   16.888   &   0.526  \\
      V6    &   RRab  &   13:13:03.9   &   18:10:19.8   &   123.7   &   0.664020(1)   &   14.4    &   16.819   &   0.522  \\
      V7    &   RRab  &   13:13:00.9   &   18:11:29.7   &   113.3   &   0.5448584(6)  &   13.5    &    \dots   &   \dots  \\
      V8    &   RRab  &   13:13:00.4   &   18:11:05.1   &   92.3    &   0.615528(1)   &   16.1    &    \dots   &   \dots  \\
      V9    &   RRab  &   13:13:00.1   &   18:09:25.0   &   81.9    &   0.6003690(7)  &   15.2    &    \dots   &   \dots  \\
      V10   &   RRab  &   13:12:45.7   &   18:10:55.6   &   143.3   &   0.6082612(7)  &   18.8    &   16.837   &   0.463  \\
      V11   &   RRabB &   13:12:45.4   &   18:09:01.9   &   156.3   &   0.629940(5)   &   12.2    &    \dots   &   \dots  \\
      V14   &   RRab  &   13:13:20.5   &   18:06:42.8   &   415.6   &   0.5454625(7)  &   14.0    &   16.880   &   0.452  \\
      V15   &   RRc   &   13:13:12.4   &   18:13:55.0   &   332.5   &   0.3086646(9)  &   21.6    &   16.894   &   0.342  \\
      V16   &   RRcB  &   13:12:46.2   &   18:06:39.2   &   247.3   &   0.3031686(7)  &   18.7    &    \dots   &   \dots  \\
      V17   &   RRc   &   13:12:40.4   &   18:11:54.1   &   236.7   &   0.381282(1)   &   27.5    &   16.848   &   0.422  \\
      V18   &   RRc   &   13:12:48.7   &   18:10:12.9   &   93.9    &   0.336054(1)   &   22.2    &    \dots   &   \dots  \\
      V19   &   RRc   &   13:13:07.0   &   18:09:26.1   &   173.2   &   0.391377(1)   &   28.1    &   16.880   &   0.437  \\
      V20   &   RRc   &   13:13:09.0   &   18:04:16.2   &   404.7   &   0.384337(1)   &   22.5    &   16.875   &   0.407  \\
      V23   &   RRc   &   13:13:02.3   &   18:08:35.9   &   137.9   &   0.365804(1)   &   25.2    &   16.825   &   0.391  \\
      V24   &   RRab  &   13:12:47.2   &   18:09:33.0   &   120.4   &   0.763198(2)   &   20.3    &    \dots   &   \dots  \\
      V25   &   RRab  &   13:13:04.4   &   18:10:37.2   &   133.5   &   0.705162(1)   &   16.7    &   16.779   &   0.555  \\
      V26   &   RRc   &   13:12:35.7   &   18:05:20.5   &   401.7   &   0.391106(1)   &   24.1    &   16.845   &   0.395  \\
      V27   &   RRab  &   13:12:41.4   &   18:07:23.9   &   257.9   &   0.671071(1)   &   14.7    &   16.856   &   0.550  \\
      V28   &   RRab  &   13:12:42.1   &   18:16:37.9   &   430.9   &   0.6327804(7)  &   16.5    &   16.875   &   0.545  \\
      V29   &   RRab  &   13:13:04.3   &   18:08:46.9   &   152.9   &   0.823243(4)   &   26.6    &   16.782   &   0.588  \\
      V31   &   RRab  &   13:12:59.6   &   18:10:04.6   &   61.9    &   0.705665(1)   &   15.2    &    \dots   &   \dots  \\
      V32   &   RRc   &   13:12:47.7   &   18:08:35.9   &   142.6   &   0.390623(2)   &   20.4    &   16.712   &   0.416  \\
      V33   &   RRab  &   13:12:43.9   &   18:10:13.2   &   162.5   &   0.6245815(8)  &   16.9    &    \dots   &   \dots  \\
      V34   &   RRc   &   13:12:45.7   &   18:06:26.1   &   262.0   &   0.289611(1)   &   21.2    &   16.928   &   0.314  \\
      V35   &   RRc   &   13:13:02.4   &   18:12:37.5   &   179.8   &   0.372666(2)   &   22.7    &   16.943   &   0.442  \\
      V36   &   RRc   &   13:13:03.3   &   18:15:10.2   &   321.8   &   0.373242(1)   &   21.5    &   16.879   &   0.414  \\
      V37   &   RRab  &   13:12:52.3   &   18:11:05.1   &   69.6    &   0.717615(1)   &   21.0    &    \dots   &   \dots  \\
      V38   &   RRab  &   13:12:57.1   &   18:07:40.5   &   151.8   &   0.705792(2)   &   21.2    &   16.749   &   0.557  \\
      V40   &   RRc   &   13:12:55.9   &   18:11:54.7   &   105.3   &   0.3147939(9)  &   19.6    &    \dots   &   \dots  \\
      V41   &   RRab  &   13:12:56.8   &   18:11:09.3   &   63.6    &   0.614438(1)   &   19.2    &    \dots   &   \dots  \\
      V42   &   RRab  &   13:12:50.6   &   18:10:19.5   &   67.4    &   0.713717(2)   &   21.8    &    \dots   &   \dots  \\
      V43   &   RRab  &   13:12:53.1   &   18:10:55.7   &   55.5    &   0.712017(2)   &   24.2    &    \dots   &   \dots  \\
      V44   &   RRc   &   13:12:51.5   &   18:09:58.2   &   54.6    &   0.375099(2)   &   23.7    &   \dots   &   \dots  \\
      V45   &   RRab  &   13:12:55.2   &   18:09:27.4   &   42.4    &   0.654950(2)   &   16.2    &   \dots   &   \dots  \\
      V46   &   RRab  &   13:12:54.6   &   18:10:36.2   &   28.1    &   0.703655(3)   &   22.0    &   \dots   &   \dots  \\
      V47   &   RRc   &   13:12:50.4   &   18:12:24.6   &   151.4   &   0.335377(1)   &   21.1    &   16.791   &   0.372  \\
      V51   &   RRc   &   13:12:57.8   &   18:10:50.7   &   54.7    &   0.355203(2)   &   20.2    &   \dots   &   \dots  \\
      V54   &   RRc   &   13:12:54.4   &   18:10:31.4   &   25.0    &   0.315122(3)   &   14.2    &   \dots   &   \dots  \\
      V55   &   RRc   &   13:12:53.6   &   18:10:39.2   &   37.7    &   0.443386(2)   &   26.4    &   \dots   &   \dots  \\
      V56   &   RRc   &   13:12:53.7   &   18:09:29.1   &   46.3    &   0.328796(2)   &   15.1    &   \dots   &   \dots  \\
      V57   &   RRabB &   13:12:55.5   &   18:09:58.0   &   12.1    &   0.568234(7)   &   11.8    &   \dots   &   \dots  \\ 
      V58   &   RRc   &   13:12:55.6   &   18:09:30.8   &   39.4    &   0.354954(3)   &   17.2    &   \dots   &   \dots  \\
      V59   &   RRc   &   13:12:56.7   &   18:09:20.4   &   53.4    &   0.303941(2)   &   19.7    &   \dots   &   \dots  \\
      V60   &   RRab  &   13:12:57.0   &   18:09:36.0   &   41.6    &   0.644755(2)   &   18.4    &   \dots   &   \dots  \\
      V62   &   RRc   &   13:12:54.0   &   18:10:30.1   &   27.0    &   0.359891(4)   &   13.8    &   \dots   &   \dots  \\
      V63   &   RRc   &   13:12:56.2   &   18:10:02.8   &   15.5    &   0.310476(4)   &   14.3    &   \dots   &   \dots  \\
      V64   &   RRc   &   13:12:52.6   &   18:10:12.5   &   38.5    &   0.319529(1)   &   19.2    &   \dots   &   \dots  \\
      V71*  &   RRc   &   13:12:54.5   &   18:09:54.3   &   18.8    &   0.304242(5)   &   14.5    &   \dots   &   \dots  \\
      V72*  &   RRc   &   13:12:55.8   &   18:09:50.6   &   20.7    &   0.254155(3)   &   13.5    &   \dots   &   \dots  \\
\hline
\end{tabular}
\\
\flushleft
{\footnotesize
 \underline{Notes:}
 \itemize
 \item The radial distance $d$ [\arcsec] is measured from the nominal cluster 
 center defined in Stetson's catalog of photometric standards
 (\citealt{standards}; $\alpha=13^h 12^m 55\fs26$; $\delta=+18\degr10\arcmin09\farcs8$ [2000.0]).\\
 \item Errors corresponding to the last digits of the periods are shown in parentheses 
 (see Sect.~\ref{gendesc} for details on the computation of the errors).\\
 \item The signal-to-noise ratio (SNR) was calculated in the $[0.0,20.0]$\,d$^{-1}$ frequency range.\\
 \item The $V$ and $V-I$ magnitude averages (i.e., the zero frequency constant in their 
 Fourier decompositions) are given in those cases in which the flux-to-magnitude conversion was reliable
 (see Sect.~\ref{method} and Fig.~\ref{plc-test}).\\
 \item Stars denote new detections, RRab, RRc and RRabB/RRcB stand for 
 fundamental, first overtone and Blazhko variables.\\
 \item Fourier decompositions of the RR Lyrae stars with magnitude-transformed
 light curves in $V$ color are available electronically online at CDS.\\
}
\end{table*}

\begin{table}
\caption{Modulation properties of the Blazhko-stars}
\label{blazhko}
\centering
\begin{tabular}{c c c c c}
\hline\hline
      ID   &  Type  & $f_{\rm m}$~[d$^{-1}$] &  $P_{\rm m}$~[d] &   $R_{{\rm m},1}$ \\
\hline
      V11  &  RRab  &       0.028            &  35              &    0.21           \\
      V16  &  RRc   &       0.017            &  58              &    0.41           \\
      V57  &  RRab  &       0.020            &  52              &    0.25           \\
\hline
\end{tabular}
\\
\flushleft
{\tiny
\underline{Note:} $R_{{\rm m},1}$ is the ratio of the amplitude of the largest 
modulation peak in the DFT spectrum and the amplitude of the $f_0$ pulsation 
component (see Table~\ref{rrlprop}), in $V$ flux.
}
\end{table}

\subsection{PLC relation}
\label{plc}

The tightest empirical relations existing for RR~Lyrae stars 
are the period$-$infrared K magnitude (PL$_{K}$) \citep{longmore} and the various 
period$-$luminosity$-$color (PLC) relations (see \citealt{kw01} 
[hereafter KW01]; \citealt{cortcat2008}; \citealt{caccat2008}). 
In addition to the theoretical interest in these relations \citep{bono2001,dcmc2004}, 
they have also important practical utilization by giving a simple 
and accurate way to determine distances. Furthermore, it is also an interesting 
question whether RRc stars (after a proper period shift) fit the empirical 
PLC relation spanned by the RRab stars. Since \object{M53} contains a considerable 
number of RRc stars, we can address this question with reasonable confidence here
(see also \citealt{cassisi2004} for concerning the same problem in the case of 
the variables in \object{M3}).
To decrease the number of parameters fitted, we use the slope of the PLC relation 
(i.e., $\log{P}$\,---\,Wesenheit index relation\footnote{By construction, the Wesenheit 
index is reddening-free in the case of ``standard'' interstellar reddening law (e.g., 
\citealt{cardelli89}). Interestingly, the best-fitting PLC relation with free-floating 
color coefficients yield values very close to those predicted by the standard extinction 
law (see \citealt{kj97}; and \citealt{udalski99}).}) derived by KW01 from globular cluster RRab variables, 
for the $V$ and $I$ magnitudes. (We note that this slope is in agreement 
with the one derived from the large -- but considerably noisier -- 
sample of LMC RRab stars by \citealt{ogle2003}.) 

\begin{figure}
\centering
\includegraphics[width=88mm]{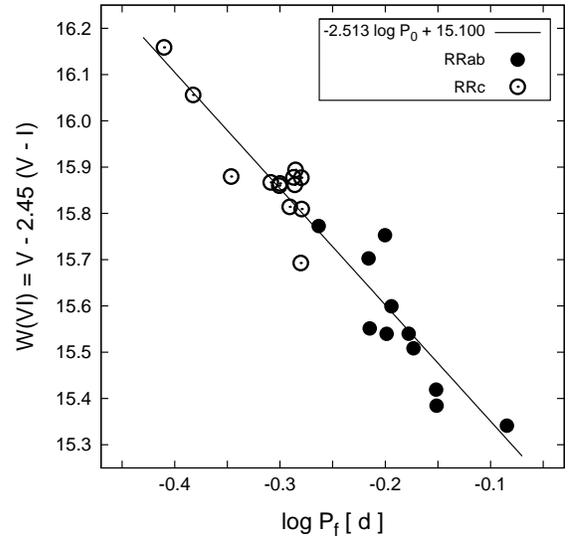}
\caption{
PLC relation for 24 RR~Lyrae stars in \object{M53}. RRc 
periods were shifted by adding $0.128$ to their $\log P_1$.
Continuous line shows the zero-point fitted regression line 
with the slope given in KW01.
}
\label{plcfig}
\end{figure}

\begin{figure}
    \centering
    \includegraphics[width=0.85\columnwidth]{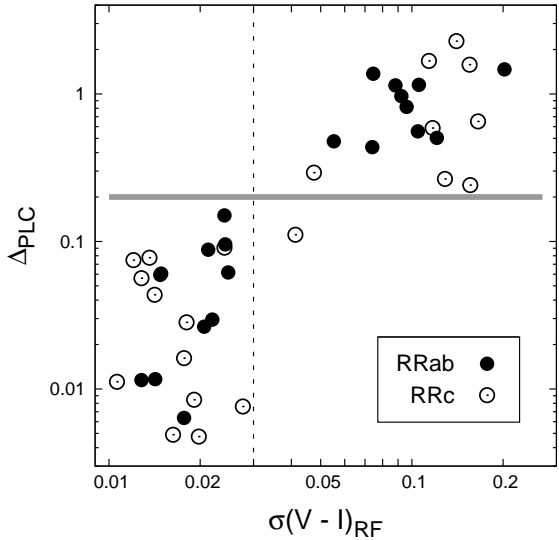}
    \caption{
Distance from the empirical PLC relation versus 
the error of the color index computed from 
the formal errors of the object images on the 
reference frames. Gray horizontal line 
shows the near $3\sigma$ limit of the scatter 
around the regression line of the empirical 
relation of KW01. Dashed vertical line indicates 
the cutoff error for the color index as explained in the text.
        }
    \label{plc-test}
\end{figure}

Figure~\ref{plcfig} shows the derived joint PLC relation. 
The periods of the RRc stars have been increased by 
$-\log{(P_1/P_0)}=0.128$ to take proper account for the overall 
period difference between the RRab and RRc stars.
The above shift corresponds to $P_1/P_0=0.745$, which is the 
average expected period ratio for the first overtone and 
fundamental mode periods for RR~Lyrae stars (see, e.g., 
\citealt{cox1983}). 
From the 54 RR~Lyrae stars with available $V,I$ photometry 
we omitted 30 stars due to their outlier status in the PLC relation
(the stars without magnitude averages in Table~\ref{rrlprop}).
We investigated the question if the omission of all these stars 
can be justified on the behavior of some other parameters 
derived during the reduction of the data. The first parameter 
to be utilized for this purpose could be the scatter of the light curve.
But while crowding introduces large errors in the zero points 
of the flux-to-magnitude conversion, this may not necessarily lead 
to the increase of the scatter of the light curve due to the effectiveness 
of the OIS method (compare variable positions in 
Table~\ref{rrlprop} and light curve scatters in Fig~\ref{rrlyrfc}). 
Therefore, we resorted the inspection of another quantity that 
might correlate with the zero point errors of the light curves. 
This quantity was the formal errors of the RF magnitudes as they 
came out from the standard method of PSF photometry. For a considerable 
number of stars, the discrepant position in the PLC relation was 
coupled with an obviously discrepant amplitude of the light curve. 
This is evidently introduced by the incorrect conversion of their 
fluxes into magnitudes (see Sect~\ref{reduction}) due to the
inaccuracy of their RF photometry. However, it is important 
to justify whether the same photometric error (and not, e.g., non-cluster 
membership) can account for the outlier status of all the omitted stars.
The correlation of the formal errors of 
$(V-I)_{\mathrm{RF}}$ with the (vertical) distance from the 
best-fitting PLC relation is exhibited in Fig.~\ref{plc-test}. 
Indeed, all omitted stars have also larger photometric formal errors. 
Therefore, we think that the data clipping used in cleaning up the PLC relation 
is justified. On the other hand, the ``good quality stars'' 
(defined as such with a formal $(V-I)_{\mathrm{RF}}$ error of 
$\lesssim 0.03$~mag, located below the gray 
line in Fig.~\ref{plc-test}) show various levels 
of agreement with the ridge line of the PLC, ranging from perfect match 
to near $3\sigma$ difference. This indicates that formal image errors 
can serve as a guide only when strong discrepancies are observed in 
applications involving photometric data derived from the lower 
quality images.
The final standard deviation of the fit is $0.063$\,mag, slightly 
worse than the one derived from several clusters by KW01.

For the zero-point calibration of this PLC relation we relied on the 
Baade-Wesselink (BW) analysis of \citet{kovacs2003}. 
From the 21 BW stars with $V-I$ color indices we omitted AV~Peg, 
because of its strongly discrepant position on the PLC relation. 
Six additional stars (X~Ari, SW~Dra, SS~Leo, V445~Oph, BB~Pup 
and W~Tuc) could also be omitted but this would not lead to a  
significant change in the zero point of the BW PLC relation 
($-1.212\pm0.037$ vs. $-1.182\pm0.026$ with the six stars omitted). 

By fixing the slope of the $\log P_0$ term as given by KW01 for 
$W(V,I)=V-2.45(V-I)$, for the 24 RRab and RRc stars of \object{M53} we get 
\begin{eqnarray}
W(V,I) & = & -2.513\log P_0 + (15.100\pm0.013)~,
\end{eqnarray}
where the error corresponds to the $1\sigma$ formal statistical error. 
Similarly, for the 20 BW stars we get
\begin{eqnarray}
W(V,I)_{\mathrm{BW}} & = & -2.513\log P_0 - (1.212\pm0.037)~.
\end{eqnarray}
Simple subtraction of the zero points of the above two equations 
yields $16.312\pm0.039$ for the dereddened distance modulus of 
the cluster. We note that this value is compatible with an \object{LMC} 
distance modulus of $18.5$\,mag (\citealt{kovacs2003}, see also 
the recent summary of \citealt{dibenedetto2008}).
Additional notes on the distance of the cluster will be given in 
Sect.~\ref{feh} in relation to the metal content and various $M_V$--[Fe/H] 
relations.

\subsection{[Fe/H] distribution}
\label{feh}

The application of the empirical formula for the computation of 
the iron abundance of RRab stars \citep[][hereafter JK96]{jk96} 
has been proven to be quite successful in the past 
(\citealt{jk96} [comparison for several GCs]; 
\citealt{szek2007} [\object{NGC\,362}]). However, there is a growing concern on 
the accuracy of this formula at the low-metallicity end 
(e.g., \citealt{nemec2004} [\object{NGC\,5053}]; \citealt{lugo2007} [\object{M15}]; 
see however \citealt{kinem2008} [Draco RRab vs. giant metallicities]).
Since \object{M53} is also a low-metallicity cluster, and the available 
light curves are of high-quality for computing precise Fourier 
decompositions, here we can examine this problem in more detail. 

In tracing back the metallicity estimates derived for this 
cluster, we found the works of \citet[][hereafter ZW84]{zw1984} 
and \citet[][hereafter S88]{suntzeff1988}. 
The metallicities presented in these papers 
are both based on calibrated spectral indices (similar to the 
$\Delta S$ index of Preston 1959 on which the JK96 formula is 
calibrated). The two studies, based on the analyses of several 
giant stars, yield very similar metallicities for \object{M53}
($-2.04\pm0.08$ by ZW84 based on the $Q_{39}$ parameter, and 
$-2.09\pm0.09$ and $-2.20\pm0.33$ by S88 based on the $m_{\rm HK}$ 
and $m_{\rm Mg}$ indices, respectively). 
High-dispersion detailed abundance analysis on a single giant 
was made by \citet{pil1983}. They obtained a value 
of $-1.9$. Additional early [Fe/H] estimates are also listed 
in their paper. Most of the values indicate a metallicity lower 
than $-1.8$. 

To further confirm the low metallicity of the giants of \object{M53}, we 
employed the photometric method of \citet{kovacs2008bvifeh} 
\citep[see also][]{dekany2008}. The method is based on the well-known 
observation that the $B-V$ color index depends fairly sensitively 
on the gravity and more importantly on the metallicity. Because 
the most relevant parameters determining the theoretical values 
of the broad-band colors are the effective temperature, gravity 
and overall metal content ([M/H], identical to [Fe/H], assuming 
solar-type heavy element distribution), we can estimate [Fe/H] by 
using $BVI$ magnitudes, assuming that we can give a reasonable 
estimate on $\log g$. For Cepheids and RR~Lyrae stars this estimate 
is very accurate, since the period is in tight correlation with 
the gravity. For giants we can get estimates on the possible ranges 
of the gravity by employing evolutionary models. We took the 
isochrones from the evolutionary models of \citet{pietr2004} 
with metallicities bracketing the estimated metallicity of $-2.0$~dex. 
Then, we checked the position of the selected giants (see Appendix~\ref{a1-rgb}) 
with available three-color photometry on the 
$\log g \rightarrow (B-V), (V-I)$ diagrams. We then took the 
corresponding $\log g$ ranges and computed individual metallicities 
from the stellar atmosphere models of \citet{cast99}. 

\begin{figure}
\centering
\includegraphics[width=88mm]{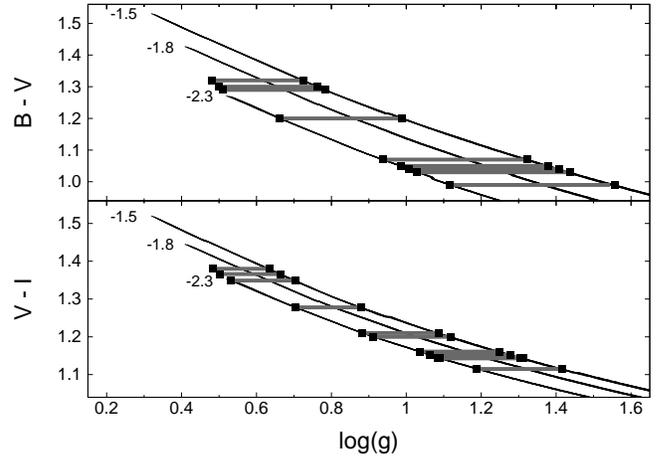}
\caption{
Estimation of the gravity of the \object{M53} giants from colors by using the evolution 
models of \citet{pietr2004}. Overall iron abundances are shown 
at the left of the isochrones spreading from 10 to 13.5 Gyr (please note that 
the age dependence is highly degenerate). Horizontal bars denote the positions 
of the giants as listed in Table~\ref{rgbprop}.
}
\label{plotisoc}
\end{figure}

The relation between $\log g$ and the color indices at various 
metallicity values and cluster ages is shown in Fig.~\ref{plotisoc}. 
It is very fortunate that there is basically no age-dependence on 
this diagram, so one can read off the corresponding $\log g$ ranges 
affected only by the adopted metallicity values. It turns out that 
in the color index -- $\log g$ space of the giants studied, there 
is a very minute dependence of the derived metallicity on the  
assumed values of $\log g$. (We note however, that for $\log g$ 
values exceeding the high-metallicity bounds by $\sim 0.3-0.5$~dex we 
could get a significant increase of up to $0.5$~dex in [Fe/H].) 
With the $\log g$ ranges confined in this way, we computed more 
accurate individual photometric abundances by matching both color 
indices simultaneously. The derived values, together with other 
important data on the stars used, are given in Table~\ref{rgbprop}.

\begin{figure}
\centering
\includegraphics[width=\columnwidth]{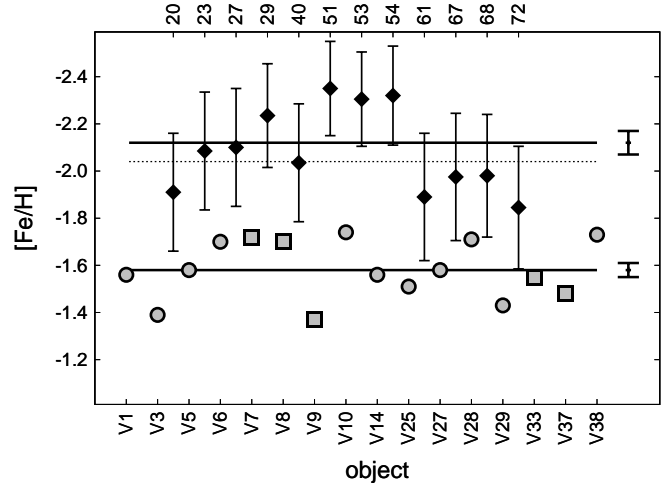}
\caption{
Photometrically derived overall iron abundances of 12 RGB stars of \object{M53} 
(black diamonds) versus iron abundances of 16 RRab stars computed from the 
$V$ light curves by the JK96 formula (gray circles [data from this paper] 
and squares [data from \citealt{kop2000}]). Horizontal lines show the corresponding 
averages (for comparison, we also show the value given by \citealt{zw1984} by 
dashed line). Formal $1\sigma$ errors are given for the means. For 
star IDs we refer to Tables~\ref{rrlprop} and \ref{rgbprop}.
}
\label{prrfeh}
\end{figure}

For comparison with the values obtained on RRab stars by using the 
empirical formula of JK96, we plotted the metallicities for the two 
groups of stars in Fig.~\ref{prrfeh}. Error bars of the individual 
RGB stars denote standard deviations, obtained by adding uncorrelated 
Gaussian random noise with $\sigma = 0.02$\,mag to their $B-V$ and $V-I$ 
color indices (the very small formal errors for the RRab stars are not shown).
In spite of the large scatter, 
it is clear that the RGB stars have significantly lower overall 
metallicities. Indeed, the average metallicities with their formal 
$1\sigma$ errors are $-2.12\pm0.05$ and $-1.58\pm0.03$, i.e., the 
difference is well above the $3\sigma$ level. We note that our 
average photometric metallicity shows fine agreement with the ones 
obtained by spectroscopic measurements, so the suspected systematic 
difference between the HB and RGB populations seems to be confirmed. 
However, we do not know at this point what is the source of this 
difference. Although we cannot exclude that there are substantial 
chemical inhomogeneities in some clusters (e.g., \object{NGC\,1851}, see 
\citealt{lee2009}), we think that it is more 
probable that the JK96 formula is biased upward at lower 
metallicities due to the low number of low-metallicity stars that 
entered in the calibration sample (see also \citealt{schcz1998}; 
\citealt{kov2002}; \citealt{nemec2004}). To check if this is indeed the case, 
we would need either precise $BVI$ photometry, or, even better, 
high-dispersion spectroscopy for the RR~Lyrae stars. Unfortunately, 
there are no such data available. We made a humble attempt to utilize 
some of the sporadic multicolor data for, e.g., \object{M15}, \object{M68}, 
and \object{M92} (\citealt{corwin2008}; \citealt{walker94}; \citealt{kj97} 
and references therein) but either the number of objects was low or 
the average magnitudes were too poor to derive consistent metallicities.

Since there are ample amount of RRc stars in M53, we decided to 
employ the formula calibrated by \citet{morgan2007} for an 
independent estimation of the metallicity of the RR~Lyrae population. 
By using their Eq.~(3) and transforming the values to the scale 
of \citet{j95}, we get an average [Fe/H] of $-1.68\pm0.08$ (error 
of the mean) for the sample of 15 stars (for two stars, data were 
taken from \citealt{kop2000}). Although this value confirms 
the one derived from the RRab population, we note the following: 
(i) the structure of the RRc light curves contain less 
distinct features than those of the RRab stars and even these are 
often washed out by noise; (ii) the calibration for the RRc stars 
is based on cluster variables that, because of the absence of 
individual metallicity estimates, are assumed to have same 
metallicity as that of the host cluster (i.e., there are $12$ 
independent metallicity values used in the regression for the 
106 variables in the calibrating sample); because of the use of 
cluster variables, the full span of metallicity range is only 
$[-1.0, -2.2]$, that is nearly a factor of two smaller than the 
one used by JK96 for the Galactic field RRab stars. We presume 
that the good overall agreement between the metallicities is 
primarily attributed to the strong dependence of [Fe/H] on the 
period for both types of stars.

We made yet another test on the possible dependence of the derived 
metallicity on the chemical composition (namely on $\alpha$-element 
enhancement, since the cluster is of low-metallicity). It turned 
out that the $\alpha$-enhanced models shift the lower boundary of 
the $\log g$ range downward by some $\sim 0.2$~dex and  
resulted in a small (i.e., $<0.1$~dex) and apparently non-systematic 
change in the metallicity [M/H]. We recall that if the solution of the 
discrepancy between the JK96 and RGB metallicities lays in the non-solar 
chemical composition, then we would expect a dramatic increase in the 
overall RGB metallicity [M/H] estimated by the above method from 
the $\alpha$-enhanced models, since then the relative abundance of iron 
(which the JK96 formula is calibrated to) would be lower in these models.

Following the recommendation of the referee, based on the above 
results we can check how the cluster's distance, as derived from the 
PLC relation, relates to the one estimated from the more traditional 
(but less accurate, due to evolutionary effects, see, e.g., 
\citealt{cassisi2004}) $M_V$--[Fe/H] relation calibrated by the same 
BW distance scale. Relying on the same BW sample as in Sect.~\ref{plc}, 
and using the same spectroscopic metallicities as in \citet{dekany2008} 
we get the following linear calibration between the absolute $V$ magnitude 
and the iron abundance:
\begin{equation}
\label{mvfeh}
M_V = (0.24 \pm 0.06)~{\rm [Fe/H]}_{\rm JK96} + (0.88 \pm 0.06)~.
\end{equation}
As expected, the standard deviation of the above fit is huge ($0.17$\,mag as 
compared to $0.063$\,mag of the PLC relation -- see Sect.~\ref{plc}).
Although many variants of $M_V$--[Fe/H] calibrations can be found throughout 
the literature, the slope of the above formula turns out to be very close to 
the value of $0.23 \pm 0.04$, favored by recent reviews in this subject 
\citep[see, e.g.,][and references therein]{catelan09}. The mean $V$ magnitude 
of M53 RR~Lyrae stars is equal to $16.85 \pm 0.01$~mag, based on the 
magnitude-transformed light curves of 24 stars (see Table~\ref{rrlprop}). 
Using the two different metallicity values derived for the RR~Lyrae and red 
giant stars, from Eq.~(\ref{mvfeh}) we get $16.35 \pm 0.11$ and $16.48 \pm 0.14$, 
respectively, for the distance modulus of the cluster. Both are in agreement 
with the PLC result in Sect.~\ref{plc} within the errors, but with an apparent 
preference for higher metallicity.

\section{Short-period variables}
\label{sxp}

\begin{figure}
\centering
\includegraphics[width=\columnwidth]{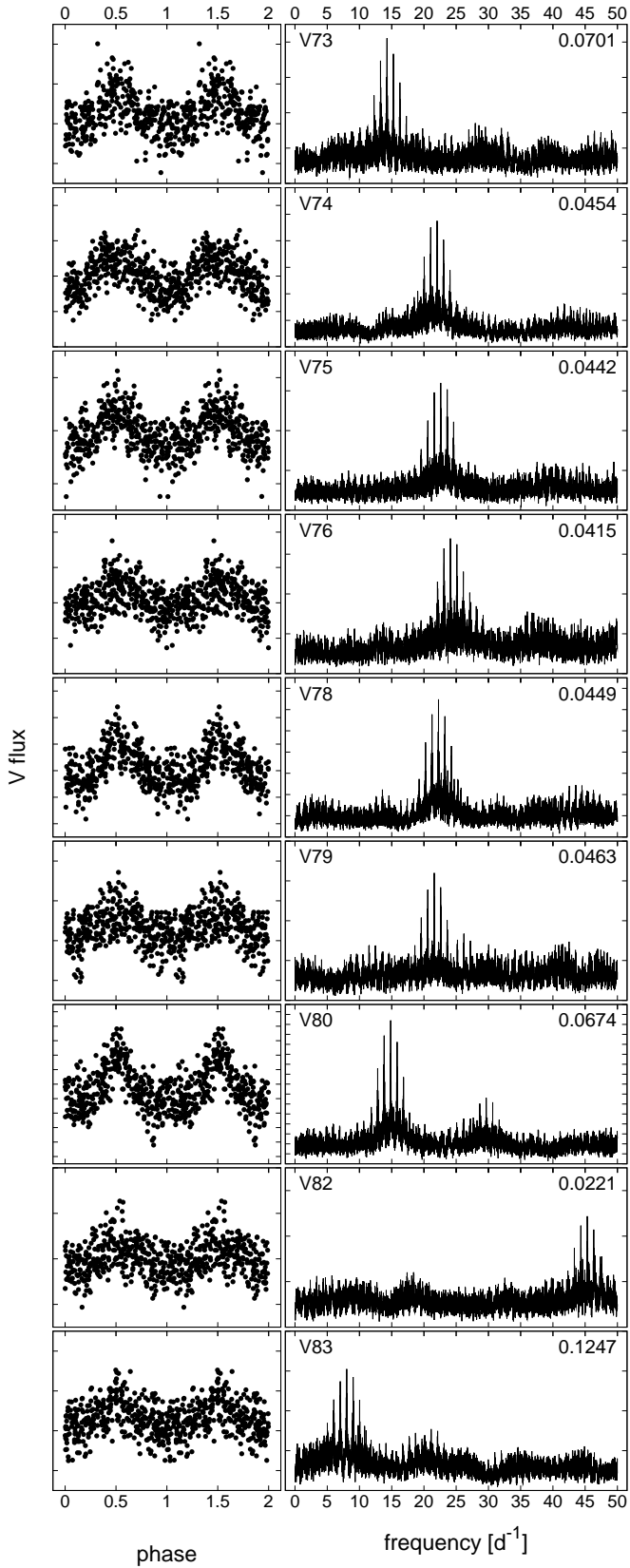}
\caption{
Left panels: TFA-reconstructed and folded light curves of the short periodic stars.
Right panels: DFT amplitude spectra of the corresponding time-series. 
The separation between ordinate ticks denotes the same flux difference in each plot.
IDs and periods are given in each case. See Table~\ref{spv-prop} for additional 
properties of these variables. 
}
\label{shortp}
\end{figure}

\begin{figure}
\centering
\includegraphics[width=\columnwidth]{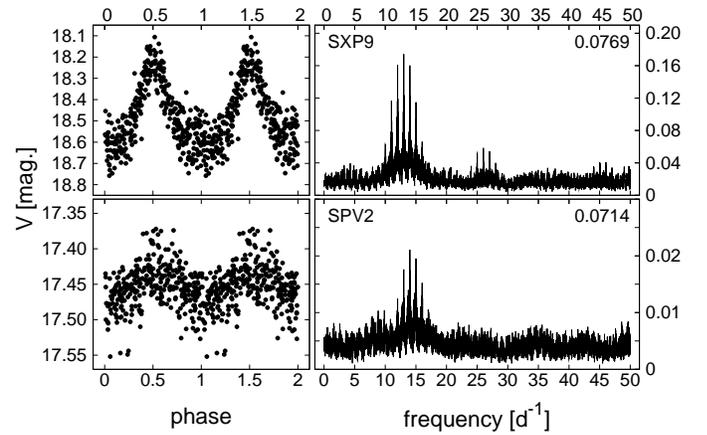}
\caption{
TFA-reconstructed light curves and DFT amplitude spectra of V77 and 
V81. Labeling is the same as in Fig.~\ref{shortp}.
}
\label{sxp9-spv2}
\end{figure}

We Fourier-analyzed the photometric time-series of all identified 
objects in the $[0,50]$\,d$^{-1}$ frequency range to search for short 
periodic variables, in particular for SX~Phoenicis and high-amplitude 
Delta Scuti stars. With the aid of TFA, we considerably increased 
the reliability of detection of short-period, low-amplitude signals 
by filtering out low-frequency trends from the data whose amplitudes 
would have overwhelmed the intrinsic light variations of these objects.

All of the short-period stars previously known in \object{M53} were discovered 
by \citet{jeon2003} who used a 1.8-m telescope and PSF photometry.
All of these objects are of SX~Phe-type located in the blue straggler 
region of the color-magnitude diagram (CMD). We successfully identified 
SXP1, SXP2, SXP6, and SXP7 among these eight objects in our data
(they are identical to our V73, V74, V75, and V76, respectively).
From the data published by \citet{jeon2003}, SXP4, SXP5, and SXP8 have too 
low amplitudes to be identified in our data. The situation is different 
for SXP3 which has a reasonably large amplitude of 0.1~mag according to 
\citet{jeon2003} but it is rather close to V66 and V67 and the limited 
resolution of our CCD chip did not enable us to separate their light 
variations.

In addition to the previously known SX~Phe stars, we discovered seven 
new objects (signed as V77 through V83) that have significant 
high-frequency signals. Except for 
two of them, these faint objects were heavily crowded or merged with 
brighter stars, therefore, we were not able to obtain accurate RF 
magnitudes for them to perform the flux-to-magnitude conversion. 
To investigate if these objects were located in the blue straggler 
star (BSS) region, i.e. they 
could be securely classified as SX~Phe stars, we sought for 
supplementary information in the literature. We cross-correlated 
the positions of the 7 new objects with the single-epoch data-base 
of \object{M53} BSSs by \citet[][via private comm.]{beccari2008} and found 
a match with sub-pixel accuracy in three cases, namely V77, V78, and 
V79, which suggests that these variables are most probably of SX~Phe type.
For V77 ($V=18.478$, $V-I=0.433$) and V81 ($V=17.454$, $V-I=0.815$), 
we were able to perform the conversion of fluxes into magnitudes. 
The average magnitudes of V77 verifies its location in the BSS region, 
while V81 is located exactly on the lower red giant branch 
\citep[see the CMD of][]{beccari2008}. We are not sure whether 
there is a blending effect of two stars in this case or V81  
(which we identified as \object{USNO-B1.0 1081-0245849}) is indeed 
a peculiar variable. We note however that there is a suspected BSS 
(\object{USNO-B1.0 1081-0245846}) near V81 which might account 
for this variability, although we did not detect the corresponding 
signal in its time-series.

The type of the remaining three new short-period variables 
(V80, V82, and V83) remains unknown, but their blue straggler 
nature cannot be ruled out (for example, we note that the previously 
known variable V73 was not identified as a BSS by \citealt{beccari2008}). 
The properties of all 11 short-period variables detected in this study 
are given in Table~\ref{spv-prop}. The light curves and amplitude 
spectra of the objects are shown in Figures~\ref{shortp}--\ref{sxp9-spv2}. 

\begin{table}
\caption{Properties of the short period stars}
\label{spv-prop}
\centering
\begin{tabular}{l c c r r r}
\hline\hline
      ID      & $\alpha$\,[hms] & $\delta$\,[dms] &  $d\,[\arcsec]$ &   $f$\,[d$^-1$]   &   SNR  \\
\hline
 V73   &   13:13:03.1   &   18:09:26.1   &   119.9   &    14.26581      &   10.2   \\
 V74   &   13:12:49.7   &   18:07:26.3   &   181.9   &    22.03949      &   13.7   \\
 V75   &   13:13:09.4   &   18:09:39.5   &   203.9   &    22.60022      &   11.8   \\
 V76   &   13:13:05.0   &   18:08:35.3   &   168.4   &    24.10224      &   10.4   \\
 V77*  &   13:13:20.8   &   18:15:34.9   &   488.3   &    13.00995      &   23.0   \\
 V78*  &   13:12:49.8   &   18:08:56.0   &   106.9   &    22.25488      &   12.7   \\
 V79*  &   13:12:46.6   &   18:11:36.9   &   150.6   &    21.59184      &    9.7   \\
\hline
 V80*  &   13:12:57.4   &   18:10:13.8   &    31.3   &    14.83029      &   14.2   \\
 V81*  &   13:13:02.7   &   18:06:29.4   &   244.5   &    14.01180      &    9.5   \\
 V82*  &   13:12:56.5   &   18:13:10.0   &   181.1   &    45.33103      &    8.4   \\
 V83*  &   13:12:50.1   &   18:07:42.9   &   164.5   &     8.01907      &    8.2   \\
\hline
\end{tabular}
\\
\flushleft
{\tiny
\underline{Notes:} New discoveries are marked with asterisks.
The signal-to-noise ratio (SNR) was calculated in the 
$[0,50]$\,d$^{-1}$ frequency range. See Table~\ref{rrlprop} for other notations.
}
\end{table}

\section{Long-period variables}
\label{lpv}

Most of the stars on the upper part of the giant branch of \object{M53} 
show some kind of low-amplitude light variations on time-scales 
of greater than $10$\,d. We identified V65, V66, V67, V69, and V70 
(see the CVSGC for their positions) among the previously known variable 
red giants, all of them showing a large power excess at very low 
frequencies ($\sim0.01$\,d$^{-1}$), implying long-term irregular 
variations.

\begin{figure*}
    \centering
    \includegraphics[width=17cm]{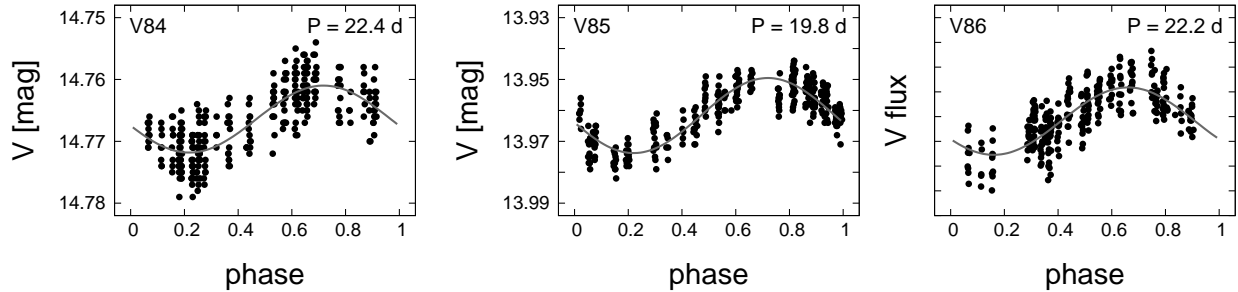}
    \caption{
$V$ light curves of the 3 new long-period red giants folded with their periods.
Star IDs, periods, and sine curves fitted to the data are also shown.
             }
    \label{lpvlc}
\end{figure*}

At the same time, we found three new long-period variables (LPVs) 
among the red giant stars (signed as V84, V85, and V86). 
The properties of these new objects are 
summarized in Table~\ref{lpv-prop} and their light curves are 
shown in Fig.~\ref{lpvlc}. Variable V86 was saturated in the $I$ band, 
therefore we give only an approximate $V$ magnitude in Table~\ref{lpv-prop} 
as a reference. V84 and V86 have very similar periods. We excluded the 
possibility of detecting the same systematics in these objects by 
using a large variety of template sets (see Sect.~\ref{tfa}) and also 
by verifying that the two light variations are not in the same phase. 

Recent investigations based mainly on OGLE-II data have showed that 
most of the pulsating red giants below the tip of the red giant branch 
(TRGB) are among the first ascent RGB stars \citep[see e.g.,][]{kb2003}. 
The majority of these stars are sitting on parallel ridges in the 
period--luminosity (PL) plane. We inspected whether the above two LPVs 
with $V-I$ color indices do match with a PL ridge on the 
$\log P$--Wesenheit index plane of \object{LMC} giants by \citet{ogle2005}, 
assuming an \object{LMC} distance modulus of $18.5$. We found that V84 is below 
the faint limit of the \object{LMC} ridges, while V85 shows a fine match with 
ridge A/R$_3$ which was associated with third-overtone pulsators by 
\citet{wood2000}.

\begin{table*}
\caption{Properties of the long period variables}
\label{lpv-prop}
\centering
\begin{tabular}{l l c c r c c c c}
\hline\hline
  ID    & $\alpha$\,[hms] & $\delta$\,[dms] &  $d\,[\arcsec]$ &   $P$\,[d]  &   SNR    &    $V$    &  $V-I$\\     
\hline
 V84  &   13:12:36.2   &  18:07:32.3    &   306.6   &    22.4     &   15.5   &   14.766  &  1.175    \\       
 V85  &   13:12:50.7   &  18:10:39.7    &    69.7   &    19.8     &   19.2   &   13.962  &  1.357    \\       
 V86  &   13:12:52.7   &  18:10:28.0    &    39.4   &    22.2     &   12.9   &   $\sim14.2$  &    \dots     \\   
\hline
\end{tabular}
\\
\flushleft
{\tiny
 \underline{Notes:} For variable V75, only instrumental $V$ magnitude average is shown 
because the star was saturated in the $I$ band. The signal-to-noise ratio (SNR) 
was calculated in the $[0,20]$\,d$^{-1}$ frequency range.
}
\end{table*}

\section{Conclusions}
\label{conclusions}

Globular cluster \object{M53} was put on our observing plan with the 
prime goal of extending the available data-base on its variable 
stars. This is the first time-series photometric study covering 
the whole cluster up to $\sim14\arcmin$ (for a comparison, the 
tidal radius of the cluster is $\sim22\arcmin$, according to 
\citealt{harris1996}).
We analyzed the whole set of $V,I$ photometric time-series 
data acquired on $3048$ objects in two seasons in 2007--2008. 
Our goal was to establish a reasonably deep variable star inventory by 
considering all possible variable types that may occur 
throughout the Hertzsprung--Russell diagram. Because of the 
limited time span of the observations, the search was limited 
to periods $\lesssim 100$~days. Similarly, the lower bound 
of the brightness level was limited by the telescope aperture 
and overall sky conditions to $V\lesssim 21$~mag, 
including the blue straggler region and ending $\sim 1$~mag below 
the main sequence turn-off point \citep[e.g.,][]{rey1998}. 
In analyzing the data we used the OIS method for image processing 
and TFA for time-series post-processing. 

Altogether $12$ new variables were identified, most of them 
are of short periodic ones (three are of SX~Phe-type). 
Interestingly, and somewhat surprisingly, except for the three 
Blazhko variables, we did not find any multiperiodic variables, 
although they could be quite abundant among RR~Lyrae and 
SX~Phe stars \citep{kov2001,gilliland1998}. 
We were also unsuccessful in finding eclipsing binaries, although some 
10\% of the blue stragglers are thought to form binary 
systems \citep{beccmmsai}. We note that in \object{NGC\,5466}, a 
similarly low metallicity cluster, \citet{ferro2008} 
identified three eclipsing variables discovered earlier 
by \citet{mateo1990}. They are all in the BSS region. 
It may well be that the relatively high scatter of our data at their 
low brightness level disabled us to detect these types of binaries.
It is interesting to note that we did not find variables 
slightly more luminous than the horizontal branch near the red giants. 
In a similar analysis of a sample of the Large Magellanic Cloud field stars, 
\citet{szkw2009} found several such stars (see also 
\citealt{ferro2008} for two such stars, one is right on 
the red giant branch of \object{NGC\,5466}). We note however, 
that we have four short periodic stars with periods reminiscent 
of the SX~Phe stars. Although, except for one object, we do not 
have color information on these, based on their very low periods 
of less than $0.13$~days, we suspect that all of them are actually 
SX~Phe variables.

In analyzing the variables identified in the present data, we 
focused on the RR~Lyrae stars. Two important parameters (average 
distance and metallicity) that are relevant for the 
physical properties of the cluster were derived from the light curves and average 
colors. We have shown that the period--luminosity--color (PLC) 
relation spanned by the RRab stars as observed in other globular 
clusters, is also very well exhibited in \object{M53}. Furthermore, it was 
also demonstrated for the first time on empirical basis that 
RRc (first overtone) RR~Lyrae stars, after proper period shift, 
follow the same linear $\log P_0$--$W(V-I)$ relation as those 
of the RRab (fundamental mode) stars. By using the Baade-Wesselink 
calibration of this relation \citep{kovacs2003}, we derived the 
distance modulus of $16.31\pm0.04$~mag for the cluster. 

By using the iron abundance--Fourier parameter relation of 
\citet{jk96}, we derived an overall iron abundance 
of $-1.58\pm 0.03$~dex for the RRab stars. This is significantly 
higher than the value of $-2.1$ computed from low-dispersion 
spectra for the red giant stars. The lower metallicity of the 
giants has been confirmed by our estimate based on $BVI$ photometry, 
stellar atmosphere and evolution models. Such a discrepancy 
of $0.3$--$0.5$~dex, seems to be present also in other 
low-metallicity clusters. For example, for the RRab stars of 
\object{M68} we find [Fe/H]$=-1.8$ \citep{jk96}, whereas 
the high-dispersion spectroscopic data on giants by \citet{lee2005} 
suggest a value of $-2.2$~dex (both values are on the 
scale of \citealt{j95}). Similarly, for the RRab stars of 
\object{NGC\,5053} \citet{nemec2004} finds [Fe/H]$=-1.7$, whereas the 
low-dispersion spectral indices of \citet{suntzeff1988} 
yield $-2.3$~dex for the giants. 

Our current understanding of this discrepancy lays in the lack 
of enough number of low-metallicity RRab stars in the original 
calibration set of \citet{jk96}. Unfortunately, 
from the works of \citet{suntzeff1994} and that of \citet{layden1994} 
there were no new efforts for obtaining metal abundances for 
large number of field stars. A reliable re-calibration of the 
formula cannot be made without a new and homogeneous sample. 
As an example for the need of new spectroscopic survey, we 
mention the case of \object{UU~Boo} \citep{jurcsik2008}, for which 
star the current Fourier formula yields [Fe/H]$=-1.17$, whereas 
the spectroscopic observations of \citet{layden1994} and \citet{kk1992} 
resulted in values of $-1.64$ and $-1.00$, respectively. 
Although other tests based on new and accurate 
measurements of non-Blazhko stars show remarkable agreement with 
earlier spectroscopic values \citep[e.g.,][]{kun2008}, the above 
example strongly supports the need of revisiting Galactic field 
RR~Lyrae stars by high-dispersion spectroscopic surveys. Within 
the same framework, a more thorough discussion of the RRab -- 
giant metallicity discrepancy would require a systematic 
high-dispersion spectroscopic study of the RR~Lyrae and other 
non-variable populations of globular clusters. Using the same 
type of measurements of all cluster members selected and 
analyzing the chemical composition in the same way 
would lead to a uniform treatment of their composition and, 
in particular, to the avoidance of the continuous confusion 
about the different zero points and scales (i.e., \citealt{zw1984} 
vs. high-dispersion values). Importance of the more 
sophisticated treatment of the metallicity issue is underlined 
by current works on detailed abundance analyses of globular cluster 
stars \citep[e.g.,][]{lee2009} and on individual field stars 
(e.g., \citealt{preston2006}; \citealt{wall09}).

\begin{acknowledgements}
We are very grateful for Dr. Giacomo Beccari for providing us 
with his data-base of \object{M53} blue straggler stars.
We are indebted to the science editor Dr. Thierry Forveille 
for the patient mediation between the referees and the authors. 
We thank the anonymous second referee for the quick and constructive
comments.
We acknowledge the support of the Hungarian Scientific Research Fund 
(OTKA, grant No. K-60750). This research has made use of the VizieR 
catalog access tool, CDS, Strasbourg, France.
\end{acknowledgements}

\bibliographystyle{aa}


\begin{appendix} 

\section{Red giant stars}
\label{a1-rgb}

\begin{table*}
\caption{Properties of the RGB stars}
\label{rgbprop}
\centering
\begin{tabular}{l l c r c c c c}
\hline\hline
   ID   & $\alpha$ [hms] & $\delta$ [dms] &$d$ [\arcsec]&    $V$     &   $B-V$   &   $V-I$  &   [Fe/H]$_{BVI}$ \\
\hline

   20   &   13:12:56.9   &   18:08:41.9   &      91.2   &   14.04    &    1.30   &   1.35  &  $-1.91\pm0.25$  \\
   23   &   13:13:03.7   &   18:09:38.7   &     123.9   &   14.07    &    1.32   &   1.38  &  $-2.09\pm0.25$  \\
   27   &   13:12:37.4   &   18:08:23.4   &     276.5   &   14.08    &    1.32   &   1.38  &  $-2.10\pm0.25$  \\
   29   &   13:12:50.4   &   18:08:53.7   &     102.6   &   14.11    &    1.29   &   1.37  &  $-2.24\pm0.22$  \\
   40   &   13:12:48.6   &   18:14:15.7   &     263.5   &   14.34    &    1.20   &   1.28  &  $-2.04\pm0.25$  \\
   51   &   13:13:17.4   &   18:14:46.0   &     419.4   &   14.56    &    1.08   &   1.21  &  $-2.35\pm0.20$  \\
   53   &   13:12:48.2   &   18:12:46.6   &     186.3   &   14.59    &    1.08   &   1.20  &  $-2.31\pm0.20$  \\
   54   &   13:13:05.3   &   18:14:50.8   &     315.7   &   14.62    &    1.07   &   1.20  &  $-2.32\pm0.21$  \\
   61   &   13:12:55.6   &   18:13:13.8   &     184.2   &   14.74    &    1.05   &   1.15  &  $-1.89\pm0.27$  \\
   67   &   13:12:38.3   &   18:11:04.7   &     248.0   &   14.80    &    0.99   &   1.12  &  $-1.98\pm0.27$  \\
   68   &   13:13:02.5   &   18:13:23.2   &     219.2   &   14.83    &    1.03   &   1.15  &  $-1.98\pm0.26$  \\
   72   &   13:12:47.9   &   18:06:32.2   &     241.5   &   14.88    &    1.04   &   1.14  &  $-1.85\pm0.26$  \\

\hline
\end{tabular}
\\
\flushleft
{\footnotesize
\underline{Note:} ID numbers and $B-V$ values were taken from the catalog of \citet{rey1998}. 
Photometric metallicities [Fe/H]$_{BVI}$ have been derived as described in Sect.~\ref{feh}.
}
\end{table*}

For the estimation of [Fe/H] using the photometric method described 
in Sect.~\ref{feh} we needed accurate $B-V$ and $V-I$ color indices 
for a number of red giant stars in \object{M53}. Since our observations covered 
only the $V$ and $I$ bands, we took the $B-V$ values from the catalog 
of \citet{rey1998}. We selected a large number of stars from this catalog 
with $V$ and $B-V$ values consistent with an RGB evolutionary state 
according to the color-magnitude diagram published by \citet{rey1998}. 
We mapped the instrumental pixel positions of the catalog onto our 
reference system by using the positions of RR~Lyrae stars common in 
Rey's catalog and our data-base. 
This transformation yielded positions with an accuracy of a few arcseconds 
which was sufficient to unambiguously identify these bright stars on our 
frames by eye. As a further selection procedure we rejected the saturated 
stars and kept only those which were well outside the core and therefore 
had reasonably low photometric errors on the RFs. After the flux-to-magnitude 
transformation, we computed the simple $V-I$ magnitude averages for these 
objects. We note that almost all RGB objects in our sample showed some 
kind of low amplitude irregular light variations on long timescales. 
Therefore we further rejected those objects for which the difference was 
too high between our $V$ magnitude and that of \citet{rey1998}. In our final 
sample of 12 RGB stars the standard deviation in the differences of the 
two $V$ magnitudes is $0.02$\,mag. Table~\ref{rgbprop} lists catalog 
IDs, coordinates, distances from the cluster center, magnitudes, and estimated 
[Fe/H] values for these 12 red giants.

\end{appendix}


\end{document}